\documentclass[twocolumn,aps,fleqn,showpacs,showkeys]{revtex4}
\usepackage{graphicx}
\usepackage{subfigure}
\usepackage{amsmath,amsfonts}
\DeclareGraphicsExtensions{.eps}
\graphicspath{{c:/users/»ç¶û/documents/matlab/}{c:/users/»ç¶û/documents/matlab/tex/}}

\begin{document}
\title{Validity of Molecular Dynamics Simulations for Soft Matter}
%
\author{Sangrak Kim}
\address{Department of Physics,\\ Kyonggi University\\ 154-42 Gwangyosanro, Youngtong-ku, Suwon 440-760, Korea}
\date{\today}
%
\begin{abstract}
In this work, we analytically examine the validity of molecular dynamics for a soft potential system by considering a simple one-dimensional system with a piecewise continuous linear repulsive potential wall having a constant slope $a$. We derive an explicit analytical expression for an inevitable energy change $\Delta E$ due to the discrete process, which is dependent on two parameters: 1) $\alpha$, which is a fraction of time step $\tau$ immediately after the collision with the potential wall, and 2) $\mu \equiv \frac{a \tau}{p_0}$, where $p_0$ is the momentum immediately before the collision. The whole space of parameters $\alpha$ and $\mu$ can be divided into an infinite number of regions, where each region creates a positive or negative energy change $\Delta E$. On the boundaries of these regions, energy does not change, \textit{i.e}, $\Delta E=0$. The envelope of $|\Delta E|$ \textit{vs.} $\mu$ shows a power law behavior $|\Delta E| \propto \mu^\beta$, with the exponent $\beta \approx 0.95$. This implies that the round-off error in energy introduced by the discreteness is nearly proportional to the discrete time step $\tau$.

\end{abstract}
%
\pacs{02.70.Ns, 02.70.Bf, 02.60.-x}
\keywords{Molecular Dynamics, Finite-difference Methods, Numerical Methods}
\maketitle
Hard, or completely impenetrable, core particles have an excluded volume effect. To simulate a hard-core system, we use event-driven methods \cite{event}-\cite{md1} to determine the time at which the hard-core particles collide. At the time of collision, the longitudinal component of the velocities is exchanged, but the transverse component of the velocities remains the same after the collision. Therefore, their total energy is completely conserved, and there is absolutely no energy drift.

In reality, atoms usually have a soft repulsive core. This can be modeled by exponential or power functional forms \cite{soft1}-\cite{soft2}. For example, noble gases such as argons can be described by the Lennard-Jones potential
\begin{equation}
    V_{LJ}(r) = 4\epsilon[(\frac{r}{\sigma})^{-12}-(\frac{r}{\sigma})^{-6}], \label{ljpot}
\end{equation}
where $\epsilon$ and $\sigma$ characterize the energy and length scales, respectively.

In molecular dynamics simulations \cite{md1}, we usually calculate the positions and velocities of the particles by solving the Newtonian equations of motion,
\begin{equation}
    m_{i}\frac{d^2\vec{r_i}}{dt^2} = \vec{F_i} (\vec{r_1}, \cdots, \vec{r_N}), i = 1, \cdots, N, \label{newtoneq}
\end{equation}
for all the particles in the system. The system governed by the Newtonian equations of motion has some interesting properties such as time reversibility, energy conservation, and so on \cite{mechanics}. Molecular dynamics is a digital computing implementation of this continuous differential equation into its corresponding discrete difference equation. Thus, keeping energy constant is a key measure in assessing the validity of molecular dynamics.

An inevitable loss of accuracy is caused by representing a derivative by its finite-difference approximation, termed a round-off error in digital computing. Many different molecular dynamics algorithms \cite{algorithms1}-\cite{algorithms3} have been proposed to solve Eq. (\ref{newtoneq}).

If the total energy of the system is different from the initial energy, the system may show very different thermodynamic and dynamic behaviors, which no longer correspond to the original objective. Therefore, it is important to know how the total energy changes with the system control parameters in the simulations.

\begin{figure}[htbp]
\begin{center}
\includegraphics[scale=.6]{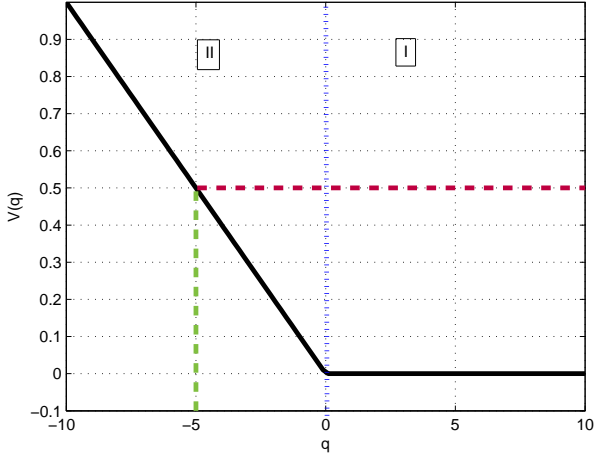}
\caption{(Color on-line) Potential model for a soft potential wall with slope $a=0.1$. If a particle has an energy $E_0 = 0.5$, continuous dynamics predicts that the particle will reflect at $q=-5.0$, but in molecular dynamics, the particle may penetrate further into the soft wall.}
\label{fig:01}
\end{center}
\end{figure}

In this article, we derive an analytical expression for energy change $\Delta E$ in a very simple system. The system is composed of a simple particle colliding with the soft potential wall in one-dimension. The soft potential wall is modeled as a piecewise linear potential, as shown in Fig. \ref{fig:01}:

\begin{equation}
$$
V(q) = \left\{ \begin{array}{rl}
 0, &\mbox{ if $q\geq 0$},  \\
  aq, &\mbox{ otherwise},
       \end{array} \label{linearpot}\right.
$$
\end{equation}
where $q$ is a coordinate of the particle and $a < 0$ is a constant that characterizes the slope of the potential. For simplicity, we take the mass of the particle as $m=1$. In region I, where $q \ge 0$, the particle moves freely with the Hamiltonian $H(p, q) = \frac{p^2}{2}$, which is quite trivial. In region II, where $q<0$, it moves with the Hamiltonian $H(p, q) = \frac{p^2}{2}+aq$.

\begin{figure}[htbp]
\begin{center}
\includegraphics[scale=.6]{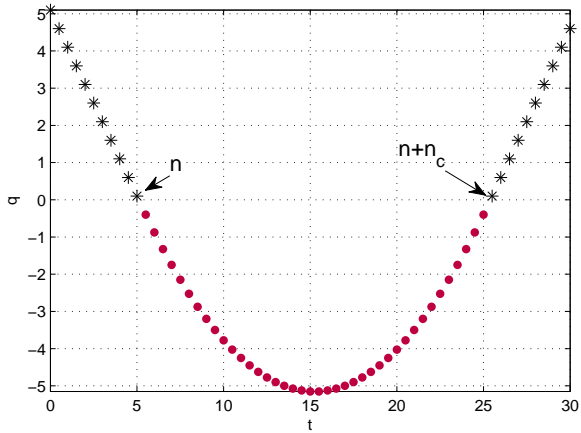}
\caption{(Color on-line) A typical particle trajectory for $a=0.1, \tau = 0.5$ starting at $q_0 = 5.1$ with $p_0 = -1.0$. The number $n$ is the number of steps immediately before the collision with the potential wall when the particle enters from region I to II. The number $n+n_c$ is the number of steps immediately after the collision with the potential wall when the particle is leaving region II. This corresponds to $\alpha=0.9$. Note that in region I, the trajectory is a straight line, denoted with a star, but in region II, it is a parabola, denoted with a filled sphere.}
\label{fig:02}
\end{center}
\end{figure}

\begin{figure}[htbp]
\begin{center}
\includegraphics[scale=.6]{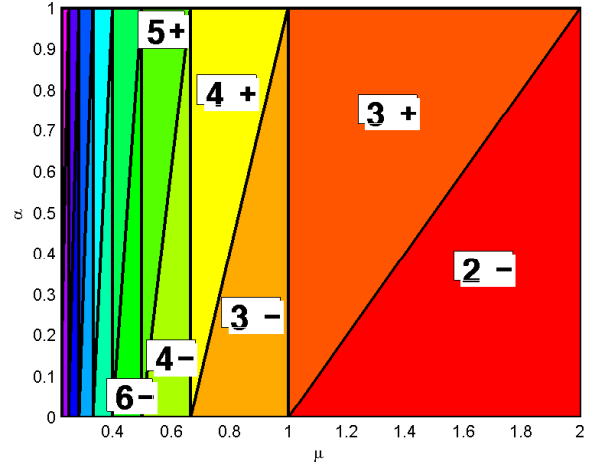}
\caption{(Color on-line) Divisions of $\alpha$ and $\mu$ space up to $n_c =10$. These divisions can continue up to $n_c \rightarrow \infty$. The cross points with abscissa are given by the formula $\mu = \frac{2}{n_c -1}, n_c = 2, \cdots, \infty$. All points on the vertical lines and diagonal lines, except $\alpha=0$ show $\Delta E = 0$. Upper right-angled triangles show $\Delta E > 0$, denoted with the $+$ symbol shown in the box. Lower right-angled triangles show $\Delta E < 0$, denoted with the $-$ symbol shown in the box. The numbers shown in the box are the number $n_c$.}
\label{fig:04}
\end{center}
\end{figure}

\begin{figure}[htbp]
\begin{center}
\includegraphics[scale=.6]{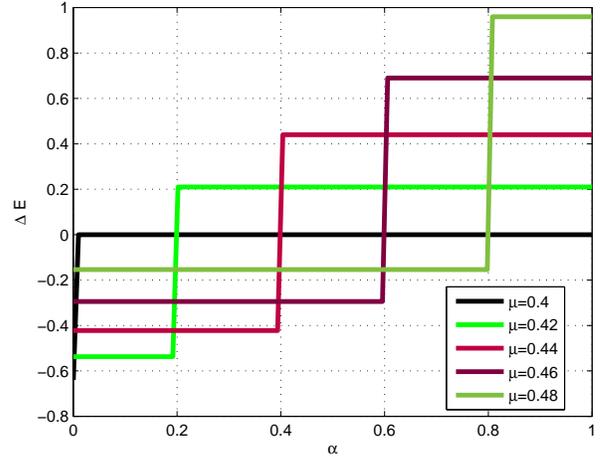}
\caption{(Color on-line) Energy change $\Delta E$  \textit{vs.} parameter $\alpha$ for different values of $\mu$. This corresponds to seeing a change along the vertical lines on the regions '4 -' and '5 +' in Fig. \ref{fig:04}.}
\label{fig:03}
\end{center}
\end{figure}

\begin{figure}[htbp]
\begin{center}
\includegraphics[scale=.53]{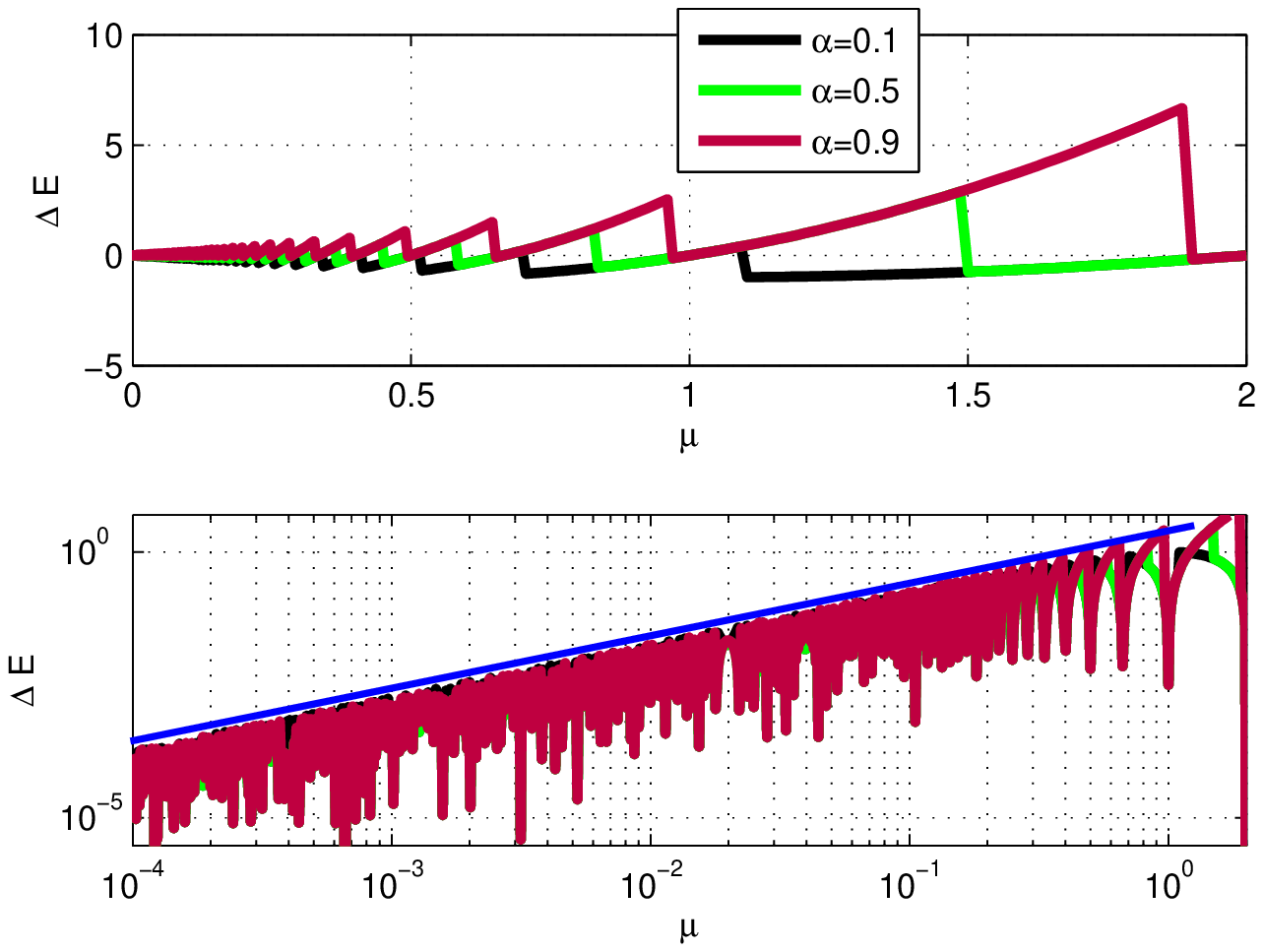}
\caption{(Color on-line) Energy change $\Delta E$ \textit{vs.} parameter $\mu$ along the horizontal lines in Fig. \ref{fig:04} for different values of parameter $\alpha$. Upper panel: $\Delta E$ \textit{vs.} $\mu$. Lower panel: $\log|\Delta E|$ \textit{vs.} $\log\mu$. The straight line shown in the lower panel has a slope of 0.95.}
\label{fig:05}
\end{center}
\end{figure}

The velocity Verlet algorithm, a symplectic algorithm, is used to solve Eq. (\ref{newtoneq}). For region I, it is written as
\begin{equation}
$$
\left\{ \begin{array}{rl}
 q_{n} ~~&=~~ q_{0} + n \tau p_{n},  \\
 p_{n} ~~&=~~ p_{0},
       \end{array} \label{verlet1}\right.
$$
\end{equation}
and for region II, it is written as
\begin{equation}
$$
\left\{ \begin{array}{rl}
 q_{n+1} ~~&=~~ q_{n} + \tau p_{n}-\frac{1}{2}a \tau^2,  \\
 p_{n+1} ~~&=~~ p_{n} - a \tau.
       \end{array} \label{verlet2}\right.
$$
\end{equation}

A typical discrete trajectory for $a=0.1, \tau = 0.5$, starting at $q_0 = 5.1, p_0 = -1.0$, is shown in Fig. \ref{fig:02}. Note that in region I, the trajectory is a straight line, denoted with a star symbol, whereas in region II, the trajectory is a parabola, denoted with a filled sphere symbol. We now examine the beginning and end of the collision with the wall. As the particle enters a line $q=0$ from region I to region II, it experiences a constant repulsive force from the wall and then finally, leaves the region II. Let $n$ denote the number of steps immediately before the collision and $n+n_c$ denote the number of steps immediately after the collision. Then, $n_c$ would be the number of steps that occur during the collision. As shown in Fig. \ref{fig:02}, when the particle enters the potential wall, it has a different value of $q_{n+1}$, depending on the previous position $q_{n}$. We parameterize this as $\alpha$, the fraction of time during which the particle moves in region II from $q_n$ to $q_{n+1}$. Thus, $\alpha =0$ corresponds to $q_{n+1} = 0$, and $\alpha =1$ corresponds to $q_{n} = 0$. The parameter $\alpha$ can take a value in (0, 1]. The case shown in Fig. \ref{fig:02} corresponds to $\alpha=0.9$. In the simulations, we actually encounter many different realizations of $\alpha$; however, unfortunately, we do not know the distribution of $\alpha$ \textit{a priori}.

We derive analytical expressions for the number of collisions $n_c$ and the relative energy change during the collision, defined by
\begin{equation}
    \Delta E \equiv \frac{{p_{n+n_c}}^2 - {p_n}^2}{{p_n}^2}. \label{F}
\end{equation}
The number of collisions $n_c$ is calculated from the inequality
\begin{equation}
    (\alpha-1) \tau p_n + n_c \tau p_n -\frac{n_c (n_c -1)}{2} a \tau^2 \ge 0, \label{mucalculated}
\end{equation}
so that $n_c$ can be obtained from
\begin{equation}
    n_c = Ceil(\frac{1+\frac{1}{2} \mu + \sqrt{{{(1+\frac{1}{2} \mu)^2}+2 (\alpha-1) \mu}}}{\mu}), \label{nccalculated}
\end{equation}
where $\mu \equiv \frac{a\tau}{p_n}$ and the function $Ceil(x)$ denotes the minimum integer larger than $x$. The quantity $a\tau$ corresponds to the magnitude of momentum generated by the force field in a time step $\tau$, and thus, parameter $\mu$ is the ratio of this momentum and the momentum of the incoming particle. The relative energy change $\Delta E$ is calculated by
\begin{equation}
    \Delta E = \xi (\xi-2), \label{Fcalculated}
\end{equation}
where $\xi \equiv \mu (n_c -1)$. Note that $n_c$ and $\Delta E$ are functions of only two dimensionless parameters, $\alpha$ and $\mu$.

Parameter $\alpha$ is in the range of $0<\alpha \le 1$, and $\mu$ is in the range of $0<\mu \le 1$ since $n_c$ is an integer greater than 1. From Eq. (\ref{nccalculated}), for a fixed value of $\mu$, the number of collisions $n_c$ changes by 1 over the whole range of $\alpha$ from 0 to 1, since $n_c (\alpha=1) - n_c (\alpha=0) =1$ for any value of $\mu$. If $\xi = \mu (n_c -1) =2$, then $\Delta E =0$ for any value of $\alpha$. Thus, vertical lines $\mu =\frac{2}{n_c -1}$ for $n_c = 2, \cdots, \infty$ divide the parameter space into an infinite number of subspaces, as shown in Fig. \ref{fig:04}. Straight lines $\alpha =(n_c -1) (\frac{1}{2}n_c \mu-1)$ for $n_c = 2, \cdots, \infty$ divide the subspaces into two regions: upper triangles show a positive change in $\Delta E$, and lower triangles show a negative change in $\Delta E$. At points on the boundaries, $\Delta E = 0$. In Fig. \ref{fig:04}, the regions are denoted with different colors and a pair of symbols including a number $n_c$ and a symbol, '+' or '-', up to $n_c = 10$.

Energy changes $\Delta E$ as a function of $\alpha$ for different values of $\mu$ are illustrated in Fig. \ref{fig:03}. These correspond to seeing a change of $\Delta E$ along the vertical lines on the regions '4 -', and '5 +' in Fig. \ref{fig:03}. For the case of $\mu = 0.4$, every point except $\alpha = 0$ gives $\Delta E =0$. For other cases, $\Delta E =0$ occurs when crossing the diagonal lines.

Fig. \ref{fig:05} shows the results of $\Delta E$ as a function of $\mu$ for different values of $\alpha$. The upper panel shows $\Delta E$ \textit{vs.} $\mu$ and the lower panel shows $\log|\Delta E|$ \textit{vs.} $\log\mu$. We find that the envelope of $\log{|\Delta E|}$ is linear with a slope of $0.95$. This exponent governs the energy fluctuation from the round-off error of a discrete time step $\tau$.

Although we have demonstrated the validity of molecular dynamics from a very simple system for soft matter, the results can be generalized to a wide range of systems. Furthermore, if we recognize many realizations of collisions, then energy fluctuations, or energy drifts in the system are coming from distributions of just two variables: 1) $\alpha$, which is related to the phase of the collision with the potential wall when the particle enters it and 2) $p_n$, which is different from collision to collision.

In summary, we considered a very simple system composed of a particle entering a linear repulsive potential wall with a constant slope $a$. By applying the velocity Verlet algorithm, the incoming momentum $p_n$ and the outgoing momentum $p_{n+n_c}$ could be derived. Further, we also derived the collision number $n_c$ and energy change $\Delta E$. These two values depended on only two parameters, $\alpha$ and $\mu$, with ranges of $0 < \alpha \le 1$ and $0 < \mu \le 1$, respectively. The parameter space created by $\alpha$ and $\mu$ could be divided into regions by the values $n_c$ and signs of $\Delta E$. An energy change of 0, \textit{i.e.}, $\Delta E =0$, was only seen on the boundaries of the regions. The envelope of $\log{|\Delta E|}$ followed a power law with an exponent of 0.95. Roughly speaking, $|\Delta E| \approx \mu^{0.95}$. Finally, any two particles with a soft repulsive potential core could be treated in the same way if we introduced the reduced coordinate system. This needs to be examined further.

\section*{Acknowledgement}
This research was supported by Basic Science Research Program through the National Research Foundation of Korea (NRF) funded by the Ministry of Education Science and Technology (2012-0002969).

%

%
\end{document}